\documentclass{elsart}

\usepackage{harvard}

\usepackage{graphicx}
\usepackage{amssymb}

\def\url#1{{\ttfamily\def\/{/\discretionary{}{}{}}#1}}

\def\HI {H\kern0.1em{\sc i}}

\begin{document}

\begin{frontmatter}
\title{Imaging \HI\ Absorption toward Symmetric Radio Galaxies -- Evidence for
a Circumnuclear Torus}

% use the thanksref command within \title, \author or \address for footnotes:
% \title{\thanksref{label1}}
% \thanks[label1]{}
% \author{\thanksref{label2}}
% \thanks[label2]{}
% \address{\thanksref{label3}}
% \thanks[label3]{}
% including your email address
% \address{\thanksref{email}}
% \thanks[email]{E-mail: }

\author{A. B. Peck\thanksref{abp}\thanksref{email}}, 
\author{G. B. Taylor\thanksref{email}}

\thanks[abp]{Present address: MPIfR, Auf dem H\"{u}gel 69, D-53121 Bonn, Germany}
\thanks[email]{E-mail: apeck@nrao.edu, gtaylor@nrao.edu}

\address{NRAO, P.O. Box O, Socorro, NM 87801}

\begin{abstract}

Recent VLBI observations have identified several compact radio sources
which have symmetric structures on parsec scales, and exhibit \HI\
absorption which appears to be associated with the active nucleus.
These sources are uniquely well suited to investigations into the
physics of the central engines, in particular to studies of the
kinematics of the gas within 100 pc of the core.  In these compact
sources, it is reasonable to assume that this circumnuclear material
is accreting onto, and ``feeding'', the central engine.
 
We present results of \HI\ imaging studies of 3 symmetric radio galaxies
which show evidence of a circumnuclear torus.

\end{abstract}

%\begin{keyword}
% keywords here, in the form keyword \sep keyword
% PACS code here, in the form \PACS code \sep code
%\PACS 
%\end{keyword}
\end{frontmatter}

% main text
\section{Introduction}
\label{intro}
A number of active galactic nuclei (AGN) have been found to exhibit HI
absorption toward the central parsecs.  Models which have been
proposed to explain this absorption suggest that disk or torus
structures exist in AGN on scales $\le$100 pc from the central engine
\cite{Con96}.  The orientation of the radio axis of the source with
respect to our line of sight determines whether or not this structure
is detectable in absorption.  Core-dominated radio sources are
oriented close to the line of sight and have relativistic jets. This
causes the approaching jet to be strongly Doppler boosted, while the
counterjet is Doppler dimmed.  Symmetric sources, on the other hand,
are oriented at larger angles to the line of sight and may have hot
spots that advance at subrelativistic velocities \cite{Ows98}.  The
continuum emission is not strongly beamed, and thus the counterjet can
contain about half of the flux density.  If the source is oriented
close enough to the plane of the sky, obscuration by a circumnuclear
torus can then be detected against the jet, counterjet, and core.  A
significant fraction of neutral atomic gas can be expected within a
range of radii determined by the midplane pressure in the
circumnuclear structure \cite{Neu95}. As predicted by this model, the
majority of detections of 21 cm atomic hydrogen seen in absorption
toward the cores of galaxies has been in Compact Symmetric Objects
(CSOs) and in extended radio galaxies which are symmetric on VLBI
scales, rather than in core-dominated radio sources \cite{vGor89}.

\section{Results}
\label{results}

\begin{figure}
% center on page
\begin{center}
% angle=-90 causes the image to be rotated counter-clockwise 90 degrees
\includegraphics*[width=11cm]{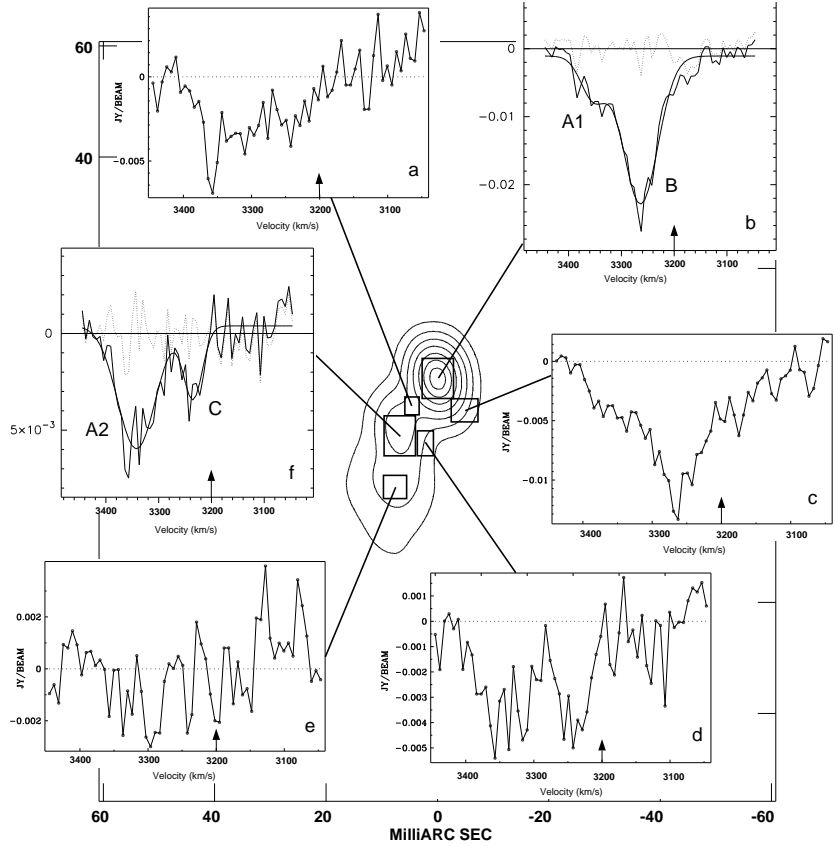}
\end{center}
\caption{\HI\ absorption profiles toward 1146+596 overlayed on 1.4 GHz
continuum contours (Peck \& Taylor 1998).  The systemic velocity is
indicated by an arrow in each panel.}
\label{pecka_fig1}
\end{figure}

\noindent{\bf 1146+596}

1146+596 is associated with the nearby ($z$=0.0107) elliptical galaxy,
NGC 3894.  Although too underluminous to be formally classified as a
CSO, 1146+596 shares the principal characteristics which lead to the
detection of \HI\ absorption in this class of objects.  Namely, it is
compact, and oriented such that a circumnuclear disk or torus should
be observable in absorption.

Strong absorption lines are detected toward both the approaching and
receding jets.  Absorption profiles integrated over 6 regions across
the source are shown in Figure \ref{pecka_fig1}.  At least two
velocity components are present in all lines of sight except toward
the SE lobe (panel \ref{pecka_fig1}e), where the signal to noise ratio
is low due to the weak continuum emission in that region.  The
strongest absorption feature is seen toward the approaching jet, shown
in panel 1b.  The optical depth of this strong feature is
$\tau$=0.12$\pm$0.01.  Although the absorption components are broader
than those found toward the center of our own Galaxy,
\citeaffixed{Lis83}{e.g.}, the blending of the lines and the low signal
to noise make it impossible to detemine conclusively whether the
absorption is associated with the AGN or due to intervening clouds in
the host galaxy.  Further observations of this source are needed.

% Note that table numbers, captions, and footnotes are left-justfied
% to the main body text in the NAR format.
\noindent{\bf PKS\,2322$-$123}

\begin{figure}
% center on page
\begin{center}
% scale= parameter can be used instead of width=
\includegraphics*[width=11cm]{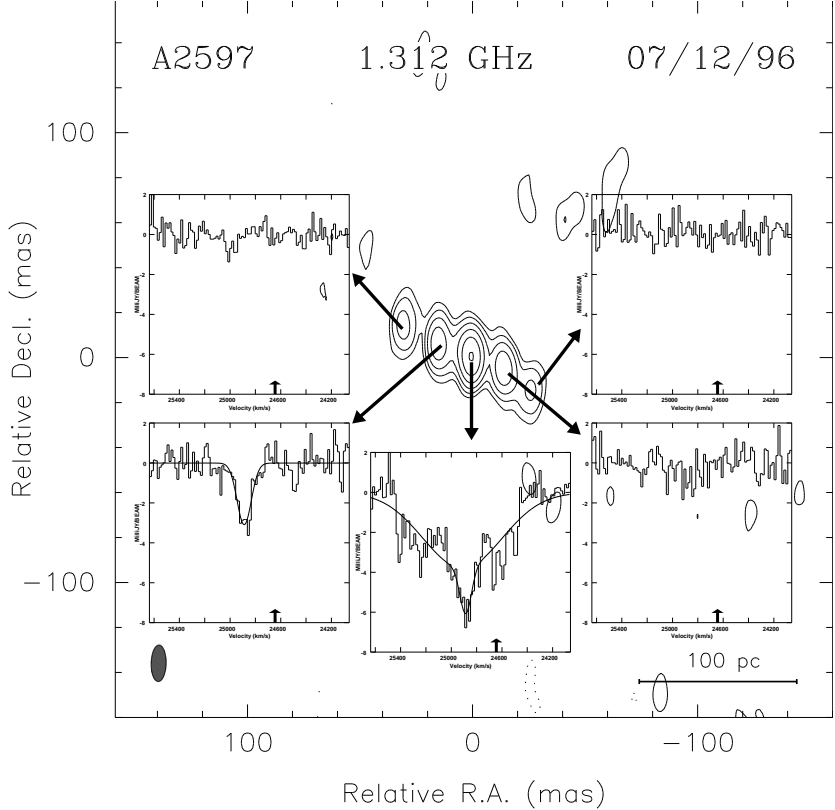}
\end{center}
% It is hard to make cites in captions come out right.  Best to do manually.
\caption{\HI\ absorption profiles toward PKS 2322-123 overlayed on 1.3
GHz continuum contours (Taylor \etal\ 1999).  The systemic velocity is
indicated by an arrow.}
\label{pecka_fig2}
\end{figure}

PKS\,2322$-$123\ is a cD galaxy at a redshift of $z$=0.082.  VLBA
observations of this source reveal straight and symmetric jets
emerging from both sides of an inverted spectrum core. \HI\ is
detected in absorption against the core and eastern jet with
substantial opacities, but is not seen towards the equally strong
western jet.  In the profiles shown in Figure \ref{pecka_fig2}, two
distinct absorption components are present, although the very broad
line (735 km s$^{-1}$ FWHM) is seen only against the core.  The
optical depth of the narrow line is $\tau$=0.63 and that of the broad
line is $\tau$=0.26.  Both lines are redshifted ($\sim$220 $\pm$ 100
km s$^{-1}$) with respect to the systemic velocity.  The fact that
both lines in PKS\,2322$-$123\ are redshifted may imply that the gas
is infalling.  This is similar to the trend found by van Gorkom \etal\
(1989) for 6 of 8 \HI\ absorption systems in a sample of nearby
ellipticals.

The most likely explanation for the observed \HI\ kinematics are an
atomic torus centered on the nucleus with considerable turbulence and
inward streaming motions. The scale height of this torus is less than
20 pc.  Unfortunately, because the radio axis of this source is so
close to the plane of the sky, we see the torus edge-on and thus have
no information about its radius.  Assuming a radius of 10 pc, the mass
required to produce the observed linewidth of 735 km s$^{-1}$ is 2
$\times 10^8$ M$_\odot$.

\noindent{\bf 1946+708}

\begin{figure}
% center on page
\begin{center}
% scale= parameter can be used instead of width=
\includegraphics*[width=11cm]{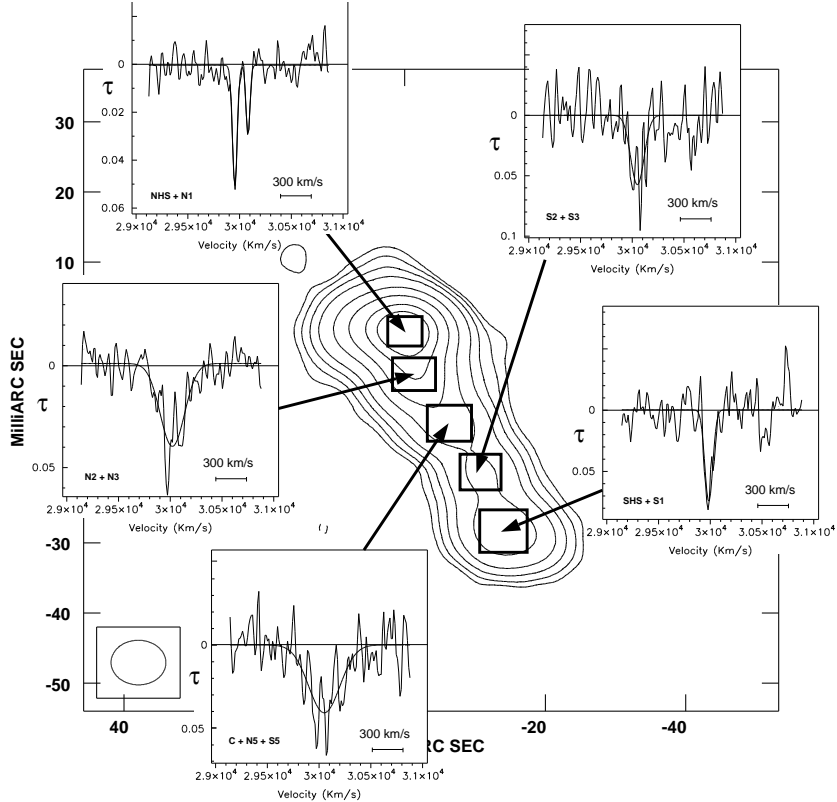}
\end{center}
% It is hard to make cites in captions come out right.  Best to do manually.
\caption{\HI\ absorption profiles toward the CSO 1946+708 overlayed on 1.3 GHz continuum contours (Peck, Taylor \& Conway 1999).}
\label{pecka_fig3}
\end{figure}

The radio source 1946+708 is a Compact Symmetric Object (CSO)
associated with a m$_{\it v}$=18 galaxy at a redshift of {\it
z}=0.101.  Figure \ref{pecka_fig3} shows the \HI\ absorption profile
toward each of 5 regions across the source.  Although there is clear
evidence of absorption toward each region, it is unclear how many
distinct components are present in each profile.  With the exception
of the profile toward the northern hot spot (NHS+N1), a single
Gaussian function has been fitted to each profile.  The systemic
velocity obtained from optical observations of both emission and
stellar absorption lines is 30279$\pm$300 km s${}^{-1}$~\cite{Sti93}.
All of the \HI\ absorption features reported here are within one
sigma of the systemic velocity.

  Narrow lines seen toward the northern hotspot which marks the
approaching jet are probably due to small clouds of \HI~ associated
with an extended ``clumpy'' torus of warm gas.  The high velocity
dispersion and column density toward the core of the source, however,
are indicative of fast moving circumnuclear gas, perhaps in a rotating
toroidal structure.  Further evidence for this region of high kinetic
energy and column density is found in the spectra of the jet
components, which indicate a region of free-free absorption along the
line of sight toward the core and inner receding jet \cite{Pec99}.
The most likely scenario to explain these phenomena consists of an
ionized region around the central engine, surrounded by an accretion
disk or torus with a radius of at least 50 pc which is comprised
primarily of atomic gas.

\section{Summary}
\label{summary}

Of the three symmetric radio sources presented here which exhibit \HI\
absorption toward the core, we find that two of them are strongly
indicative of a circumnuclear torus.  The third requires further
investigation.  All three sources, however, yield strikingly different
absorption profiles than those found in nearby asymmetric radio
sources.  In Centaurus A, for example, three very strong absorption
features are found.  The optical depths of the three lines vary little
across the continuum source, and the FWHM linewidths are between 7--20
km s$^{-1}$~ \cite{vdH83,Pec99b}.  Comparison between these two types
of absorption systems is discussed further in \citeasnoun{Pec99b}.

In the symmetric sources shown here, in which the lines are very broad
and the highest column densities are centered on the cores, it is
reasonable to assume that this circumnuclear material is accreting
onto, and ``feeding'', the central engine, and that this process will
lead to their eventual evolution into much larger FR II sources
\cite{Rea96b,Fan95}. This torus is also expected to obscure the
central parsecs of sources which happen to be oriented at large angles
to the line of sight, resulting in the observed differences between
the Type I (quasar, Seyfert 1) and Type II (radio galaxy, Seyfert 2)
objects.

% The phrase \cite{Bai92} produces (Bailyn 1992).
% In the phrase \citeasnoun{Bai95} Bailyn et al. (1995) appear as a noun.
% Affixes (e.g. Barnes et al. 1976) are produced by the phrase
% \citeaffixed{Barnes et al. 1976}{e.g.}.
% Other options of the harvard package, e.g. \citeyear, are not
% reproduced in New Astronomy.

\end{document}